\documentclass[preprint,12pt]{elsarticle}
\usepackage{tabularx}
\usepackage{lscape} 
\usepackage{adjustbox}
\usepackage{csquotes}
\usepackage{amsmath}
\usepackage{mathtools}
\usepackage{amssymb}
\begin{document}

\begin{frontmatter}



\title{Machine Learning (ML)-Centric Resource Management in Cloud Computing: A Review and Future Directions}


\author[inst1]{Tahseen Khan}

\affiliation[inst1]{organization={University of Electronic Science and Technology of China},%Department and Organization
            city={Chengdu},
            country={China}}

\author[inst1]{Wenhong Tian}
\author[inst2]{Rajkumar Buyya}

\affiliation[inst2]{organization={University of Melbourne},%Department and Organization
             city={Melbourne},
            country={Australia}}

\begin{abstract}
Cloud computing has rapidly emerged as model for delivering
Internet-based utility computing services. In cloud computing, Infrastructure as a Service (IaaS) is one of the most important and rapidly growing fields. Cloud providers provide users/machines resources such as virtual machines, raw (block) storage, firewalls, load balancers, and network devices in this service model. One of the most important aspects of cloud computing for IaaS is resource management. Scalability, quality of service, optimum utility, reduced overheads, increased throughput, reduced latency, specialised environment, cost effectiveness, and a streamlined interface are some of the advantages of resource management for IaaS in cloud computing. Traditionally, resource management has been done through static policies, which impose certain limitations in various dynamic scenarios, prompting cloud service providers to adopt data-driven, machine-learning-based approaches. Machine learning is being used to handle a variety of resource management tasks, including workload estimation, task scheduling, VM consolidation, resource optimization, and energy optimization, among others. This paper provides a detailed review of challenges in ML-based resource management in current research, as well as current approaches to resolve these challenges, as well as their advantages and limitations. Finally, we propose potential future research directions based on identified challenges and limitations in current research.
\end{abstract}


\end{frontmatter}


\section{Introduction}
\label{Sec:intro}
\begin{figure*}[t]
\begin{center}
  \includegraphics[width=.65\textwidth]{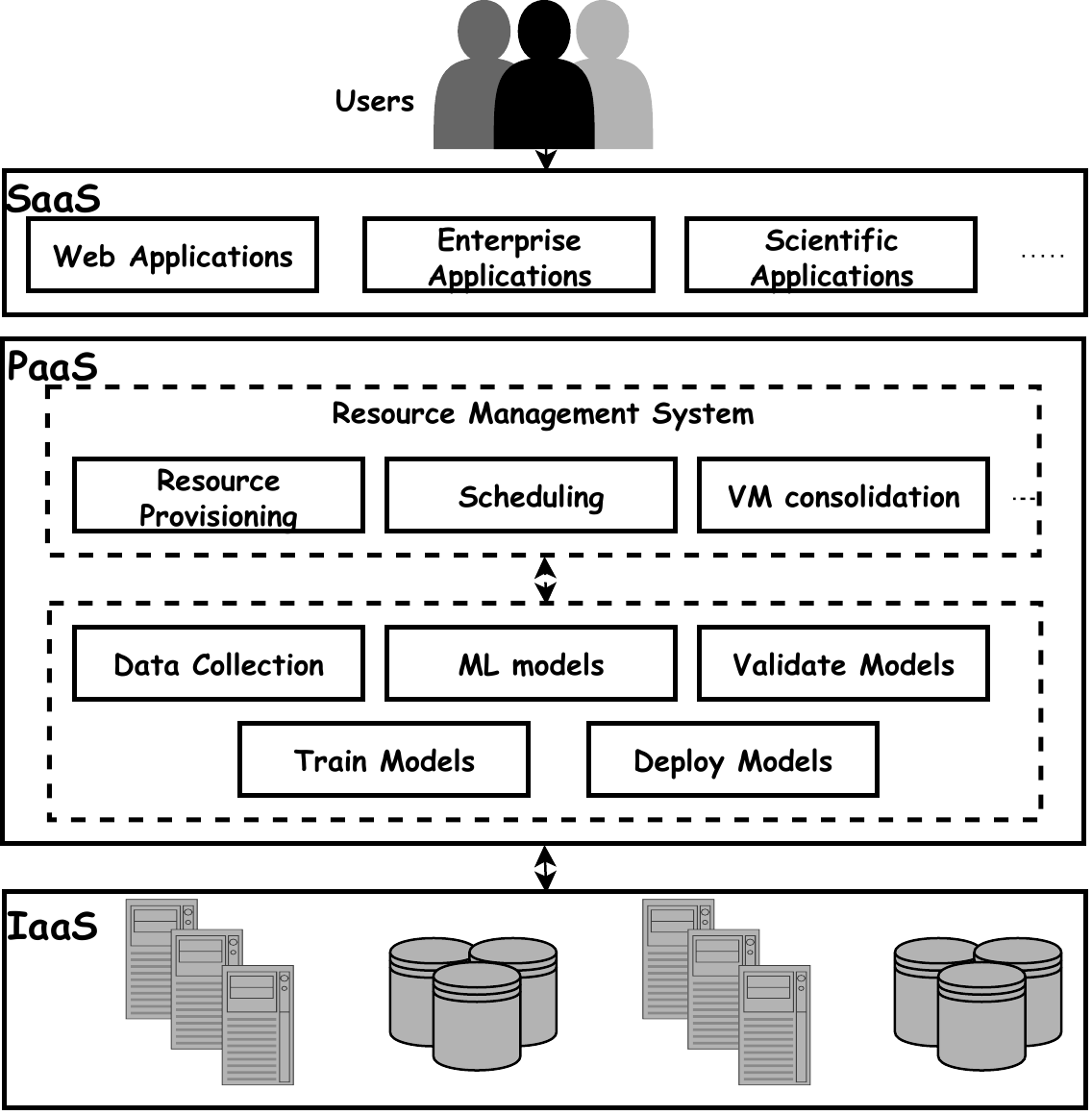}
\caption{Components of Cloud computing Paradigm using Machine Learning}
\label{fig:1} 
\end{center}
\end{figure*}
Cloud computing has created an environment in which consumers use software and IT infrastructure, paving the way toward the emergence of computing as a fifth utility \cite{10.1145/3241737}. Resource management in data centres remains a nontrivial issue in cloud computing, and it is directly dependent on the application workload. Applications were connected to specific physical servers in conventional cloud computing environments such as data centres, so these servers were often overprovisioned to handle issues related to maximum workload \cite{xu2017survey}. As a result of the wasted resources and floor space, the data centre was expensive to operate in terms of resource management. Virtualization technology, on the other hand, has proven that it can make data centres easier to handle. This technology offers a variety of benefits, including server consolidation and higher server utilisation. Large IT giants like Google, Microsoft, and Amazon have massive data centres with complicated resource management. Servers, virtual machines (VMs), and various management roles are all part of the resource management of these massive data centres, according to \cite{bianchini2020toward}. A server host is allocated multiple VMs with varying workload types and amounts in these data centres. This variable and unpredictable workload may result in a server being over-utilized and underutilised, resulting in an imbalance in resource utilisation assigned to VMs on a specific hosting server. This could lead to issues including inconsistent quality of service (QoS), unbalanced energy use, and service level agreements (SLA) violations, according to \cite{singh2019secure}. According to a survey on unbalanced workload, the average CPU and memory utilisation was 17.76\% and 77.93\%, respectively, and a similar study in the Google data centre found that the CPU and memory utilisation of a Google cluster could not exceed 60\% and 50\%, respectively \cite{kumar2020self}. As a consequence of the imbalanced workload, a data center's productivity suffers, resulting in increased energy consumption. It is proportional to the data center's operational costs and financial loss. This excessive energy consumption has a direct impact on carbon footprints, which should be reduced because an ideal machine absorbs more than half of the maximum energy consumption \cite{barroso2013datacenter}. According to an EIA (Energy Information Administration) survey, data centres consumed around 35 Twh (Tera Watt hour) of energy in 2015, and this figure is expected to rise to 95 Twh by 2040.

The resource use can be balanced by reducing the number of active servers; thus, the optimal mapping between VMs and servers must be discovered \cite{li2013energy}. This is a challenging and NP-complete problem class. As a result, an intelligent resource management strategy is needed to meet QoS requirements while also increasing data centre benefit \cite{kumar2020cloud}. The intelligent mechanisms will generate future insights, which can aid applications in mapping to machines with better resource utilisation \cite{kumar2020ensemble}. However, the nonlinear and variable behaviour of workloads for VMs creates a significant challenge when estimating future insights. However, this future insight can be obtained using two different approaches: historical workload based prediction methods, which generate insight by learning trends from historical workload data, and homeostatic based prediction methods, which provide an upcoming future workload insight by subtracting the previous workload from the current workload \cite{kumar2018workload}. Furthermore, the previous workload's mean may be static or dynamic. Both methods have advantages and disadvantages, but historical-based forecasts are considered simpler and are well-known in this field.

Thus, by conducting effective and intelligent resource provisioning, intelligent resource management will play a critical role in optimising the data center's SLA, energy usage, and operating costs. Resource management in data centres encompasses a variety of activities, including resource provisioning, reporting, workload scheduling, and a variety of other functions \cite{ilager2020artificial}. Many of these activities revolve around resource provisioning. The aim of resource provisioning is to assign cloud resources to VMs based on end-user requests while maintaining a minimum of SLA violations, such as availability, reliability, response time limit, and cost limit \cite{shahidinejad2020resource}. It should assign resources in accordance with end-user demands and prevent over or under provisioning, such as allocating more or less resources to VMs. This resource allocation technique can be carried out in two ways: proactive and reactive. In proactive approaches, resource provisioning is focused on workload prior prediction, which is estimated by learning trends from historical workload, while reactive approaches are carried out after resource demand arrives. As a result, it's inferred that historical-based prediction methods' expertise can be effectively incorporated in proactive approaches to provide intelligent dynamic resource scaling, which contributes to intelligent dynamic resource management. In addition, other functions, such as VM consolidation, task scheduling, and thermal management, can be performed based on forecasts to optimise resource utilisation, energy consumption, and increase QoS. Machine learning (ML) techniques are widely used in a variety of fields, including computer vision, pattern recognition, and bioinformatics. Large-scale computing systems have benefited from the advancement of machine learning algorithms \cite{mao2019learning}. Google recently released a report detailing their efforts to optimise electricity, reduce costs, and improve efficiency \cite{jeff2018ml}. ML has drawn attention to dynamic resource scaling by providing data-driven methods for future insights, which is regarded as a promising approach for predicting workload quickly and accurately. 

As a result, this article focuses on the review based on challenges discovered in state-of-the-art research in resource management by using ML algorithms including various resource management tasks such as provisioning, VM consolidation, thermal prediction, and other management approaches. Then we'll talk about identified the advantages and limitations of various state-of-the-art research studies in resource management that use machine learning algorithms. We will also discuss about the experimental settings along with used data sets and performance improvements. Finally, we propose future research directions based on identified challenges and limitations in current research. Fig \ref{fig:1} depicts the cloud computing components while using machine learning.

\subsection{Motivation of Research}
In cloud operations, resource management is a difficult task because multi-tenant end-users demand nonlinear workloads, which can lead to many over- and underutilised servers. It has a direct effect on whether electricity is over- or under-utilized, resulting in a high operating cost. As a result, intelligent resource management can benefit from a prior estimate of workload based on historical data. Static policies are often used in cloud computing systems to manage resources, and they have two flows: they are based on a static threshold value that is adjusted in offline mode, and they appear to require reactive behaviour, which may result in excessive overheads and delay customer responses.These strategies fail in a dynamic context, for example, when load reaches the static threshold and rapidly drops, indicating that VM migration is unnecessary in the case of VM consolidation. Furthermore, they are unable to capture the dynamics of technology and workload in complex dynamic environments (such as Cloud and Edge) and therefore fail to move through \cite{ilager2020artificial}.
To address these disadvantages, machine learning has supplanted static heuristics with dynamic heuristics that adapt to the real production workload. \cite{yadwadkar2018machine,mao2016resource}. Predictive management is made possible by machine learning techniques, which provide future insight based on historical data. As a result, A data-driven Machine Learning (ML) model in an ML-centric RMS can forecast future workload demand and control auto-scaling of resources accordingly.
Such strategies are extremely beneficial for both consumers and service providers who want to improve their QoS and keep their competitive edge in the market. For cloud resource management, ML has been shown to make more reliable predictions than more conventional approaches, such as time-series analysis \cite{cao2018load,chen2018modeling} 

Several ML algorithms have been developed to predict prior workload for intelligent resource management. Furthermore, a number of IT behemoths have begun to investigate machine learning-based resource management in production \cite{cortez2017resource, gao2014machine}. Google optimises fan speeds and other energy kobs using a neural network \cite{gao2014machine}. Microsoft Azure makes use of a framework resource central to provide online forecasts of different workloads using various ML Gradient Boosting Trees \cite{bianchini2020toward}. Despite these previous attempts and opportunities, the best way to incorporate machine learning into cloud resource management is currently uncertain. As a result, it has become critical to present research that addresses current challenges and suggests potential future research directions while also highlighting the benefits and limitations of current research.

\subsection{Our Contributions}
The following are the main contributions of our work:
\begin{itemize}
    \item We present a review of ML-based resource management approaches in cloud computing based on identified challenges in the state-of-art research.
    \item We identify the advantages and drawbacks of these methods, as well as their experimental configuration, data sets used, and performance improvements.
    \item We propose potential future research directions based on identified challenges and limitations in the state-of-art research to strengthen the resource management
\end{itemize}

\subsection{Related Surveys}
A few studies have been published on machine learning-based resource management in cloud computing. \cite{sun2016optimizing} provided a detailed survey of the most important research activities on data centre resource management with the aim of improving resource usage. After that, the article summarises two major components of the resource management platform and addresses the benefits of predicting workload accurately in resource management. \cite{manvi2014resource} focused on resource provisioning, resource allocation, resource mapping, and resource adaptation, among other essential resource management techniques. \cite{zhang2016resource} surveyed the state of the algorithms, organised them into categories, and addressed closely related topics such as virtual machine migration, forecast methods, stability, and availability.

These articles do not go into great detail about machine learning-based resource management, nor do they go into great detail about the challenges and issues that exist in the existing state-of-the-art and future research directions.  As a result, it is now important to present a thorough survey that addresses various machine learning algorithms used in the resource management scenario for a data centre, as well as their shortcomings, challenges, and potential directions, as per our vision. Hence, this article can help researchers evaluate the current machine learning scenarios in cloud resource management and their shortcomings before moving forward with their new ideas in this direction.

\subsection{Article Structure}
The remaining sections of the paper are organised as follows: The background details and definitions for cloud computing components and machine learning are given in Section 2. Section 3 discusses the challenges of machine learning-based resource management in cloud computing systems, as well as the benefits and drawbacks of current research. Section 4 proposes future research directions based on the challenges and limitations pointed out in state-of-the-art research, and Section 5 concludes the paper.

\section{BACKGROUND AND TERMINOLOGIES}
\subsection{Cloud Computing}
Cloud computing refers to the provisioning of resources over the Internet, such as memory, CPU, bandwidth, disc, and applications/services. The National Institute of Standards and Technology (NIST) \cite{nistcloud} states that \enquote{Cloud computing is a model for providing on-demand network access to a common pool of configurable computing resources (e.g., networks, servers, storage, software, and services) that can be quickly provisioned and released with minimal management effort or service provider involvement. There are five core features, three service models, and four deployment options in this cloud model}. Based on the literature, two more characteristics have been included.

This computing model uses a client-server architecture to allow for centralised application deployment and computation offloading. Cloud computing is cost-effective in application delivery and maintenance on both the client and server sides, as well as flexible in resource provisioning and detaching services from related technologies. Cloud computing and its supporting technology have been investigated for years, and many advanced computing systems have been released to the market, including Alibaba Cloud, Microsoft Azure, Adobe Creative Cloud, ServerSpace, Amazon Web Services (AWS), and Oracle Cloud.

\subsection{Core features of cloud computing
}
\begin{itemize}
    \item On-demand self-service: A client can query one or more services as needed and pay using a "pay-and-go" system without interacting with living beings via an online control center.
    \item Broad network access: Resources and services in different cloud provider areas can be accessed from a number of locations and provisioned by incompatible thin and thick clients using standard mechanisms. This trait is often referred to as \enquote{easy-to-access standardised mechanisms} and \enquote{global reach capability} \cite{hamdaqa2012cloud,yakimenko2009mobile}.
    \item Resource pooling: It offers a set of resources that act as if they were one blended resource \cite{wischik2008resource}. In other words, the client is not aware of the location of the provided services and is not expected to be. This strategy enables vendors to dynamically include a variety of real or virtual services in the cloud.
    \item Rapid elasticity: Elasticity is just another word for scalability; it refers to the ability to scale resources up or down as required. Clients can demand as many services and resources as they want at any time. Amazon, a well-known cloud service provider, named one of its most popular and commonly used services the Elastic Compute Cloud because of this consistency \cite{amazon2010amazon}.
    \item Measured service: Various facets of the cloud should be automatically controlled, monitored, optimised, and documented at several abstract levels for both vendors and customers.
    \item Multi-Tenacity: The Cloud Security Alliance proposes this idea as the fifth cloud characteristic. Multi-tenacity implies that models for policy-driven compliance, segmentation, separation, governance, service levels, and chargeback/billing for various customer categories are needed \cite{espadas2013tenant}.
    \item Auditability and certifiability: It is important that services plan logs and trails in order to assess the degree to which laws and policies are followed \cite{hamdaqa2012cloud}.
\end{itemize}

\subsection{Cloud computing service models}
\begin{itemize}
    \item Software as a Service (SaaS) \cite{piraghaj2017survey}:Using this service model, a client can access the service provider Cloud-hosted applications. Web portals are used to access applications. Since providers have access to the applications, this model has made production and testing easier for them.
    \item Platform as a Service (PaaS) \cite{jula2014cloud}: In this service model, the service provider provides basic requirements including network, servers, and operating system to enable the client to build acquired applications and manage their configuration settings.
    \item Infrastructure as a Service (IaaS) \cite{whaiduzzaman2014survey}: The user has created all of the necessary applications and only requires a simple infrastructure. Vendors may include processors, networks, and storage as facilities with customer provisions in such cases.
\end{itemize}

\subsection{Deployment models for cloud computing}
\begin{itemize}
    \item Public cloud \cite{toosi2014interconnected}: This is the most popular cloud computing model, in which the cloud owner, in the majority of cases, provides public services over the Internet based on predetermined rules, regulations, and a business model. With a significant number of commonly used resource base, providers can provide consumers with a range of choices for choosing appropriate resources while maintaining QoS.
    \item Private cloud \cite{jadeja2012cloud}: A private cloud is created and configured to provide a company or institute with the majority of the advantages of a public cloud. Setting up such a system would result in less security problems due to the use of corporate firewalls.The high costs of establishing a private cloud are a fatal flaw because the business that manages it is accountable for all facets of the scheme.
    \item Community cloud \cite{dillon2010cloud}: A variety of organisations form a group and share cloud computing with their community members' customers based on common criteria, concerns, and policies. The required cloud computing infrastructure can be provided by a third-party service provider or a group of community members. The most important benefits of a community cloud are cost savings and cost sharing among community members, as well as high protection.
    \item Hybrid cloud \cite{tuli2020shared}: Combining two or more independent public, private, or community clouds resulted in the creation of a new cloud model known as hybrid cloud, in which constituent services and infrastructure maintain their special features while also requiring standardised or agreed-upon functionalities to enable them to communicate in terms of application and data interoperability and portability.
\end{itemize}

\subsection{Machine Learning}
\begin{figure*}[t]
\begin{center}
  \includegraphics[width=\textwidth]{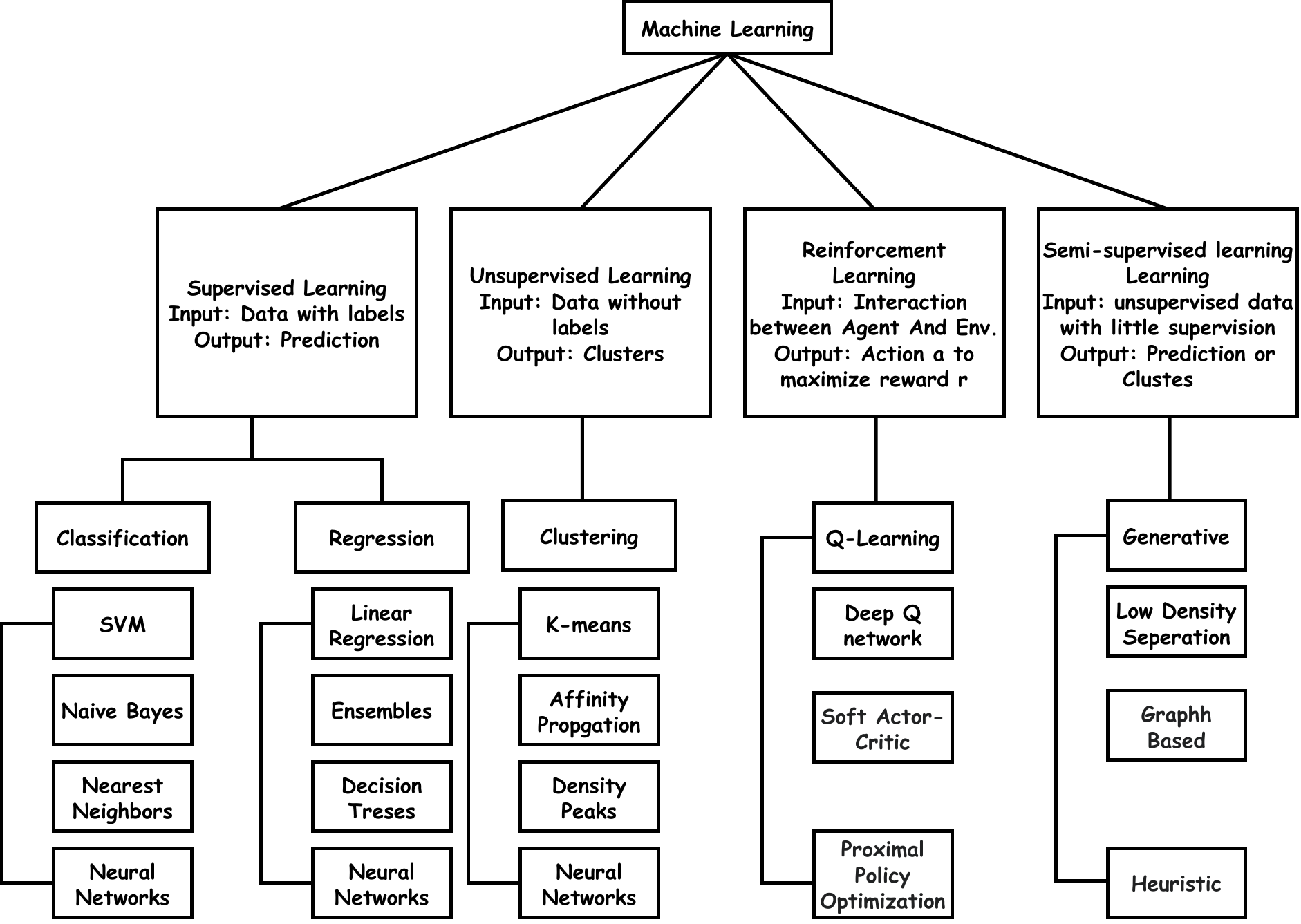}
\caption{Taxonomy of machine learning}
\label{fig:1} 
\end{center}
\end{figure*}
The study of training machines to make predictions or recognise items without being explicitly programmed to do so is known as machine learning \cite{jordan2015machine}. One of its basic assumptions is that using training data and statistical techniques, it is possible to construct algorithms that can predict potential, previously unseen values. Machine learning has come a long way in the last two decades, from a research project to a widely used commercial technology. Machine learning has emerged as the preferred tool for designing functional apps for computer vision \cite{janai2020computer}, speech recognition \cite{deng2013machine}, natural language processing \cite{olsson2009literature}, robot control \cite{chin2020machine}, self-driving cars \cite{stilgoe2018machine}, effective web search \cite{bhatia2008information}, purchase recommendations \cite{hastie2009elements} and other applications in the field of artificial intelligence (AI). Many AI system developers now understand that, for many applications, training a system by showing it examples of desired input-output actions is much simpler than programming it manually by predicting the desired answer for all possible inputs. This success is primarily owing to the accessibility of massive data and increased efficiency in the processing power of servers and GPUs \cite{goodfellow2016deep}. Based on the modelling objective and the problem at hand, machine-learning algorithms are categorised as supervised learning, semisupervised learning (SSL), unsupervised learning, and Reinforcement Learning (RL). Unsupervised learning is categorised as clustering and dimension reduction \cite{hartigan1979ak,guha2000rock,ding2002adaptive}, among other things, while supervised learning is categorised as the classification problem (e.g., sentence classification \cite{kim-2014-convolutional,yin-schutze-2015-multichannel}, image classification \cite{yang2009linear,bazi2009gaussian,ciregan2012multi}, etc.) and regression problem.

\begin{itemize}
    \item Supervised Learning \cite{sen2020supervised}: Every data sample in supervised learning is made up of several input features and a name. The learning process is designed to get as close as possible to a mapping function that links the features to the label. Following that, the mapping function can be used to make predictions of the label for the data given new input features. This is the most widely used machine learning scheme, and it's been used for a lot of things. The classification task, which involves classifying an object based on its characteristics, such as classifying mobile by its brand name and specifications, is an example of supervised learning. This is a regression task if the supervised learning task is to forecast a continuous variable like stock pricing. As shown in Fig. 1b, we can further categorise supervised learning based on the model form.
    \item Unsupervised Learning \cite{celebi2016unsupervised}: Unsupervised learning, in comparison to supervised learning, is when we only have input features but no names to go with them. As a result, the purpose of unsupervised learning is to learn the data distribution and demonstrate how the data points vary from one another. The clustering problem, which is to discover data groupings, such as grouping VMs based on their resource use patterns, is a good example of unsupervised learning.
    \item Semi-supervised learning \cite{van2020survey}: It is a branch of machine learning that attempts to integrate these two activities. SSL algorithms usually try to increase efficiency in one of these two tasks by incorporating knowledge from the other. When dealing with a classification problem, for example, additional data points with unknown labels may be used to help in the classification process. On the other hand, knowing that some data points belong to the same class will help with the learning process for clustering methods.
    \item Reinforcement Learning \cite{kober2013reinforcement}: In several ways, RL varies from supervised and unsupervised learning. It is not necessary to use labelled input/output pairs or explicit correction on sub-optimal options when using reinforcement learning to train an agent. Instead, the agent attempts to find an equilibrium between exploration and exploitation by interacting with the environment. The translator rewards the agent for successful decisions or behaviour. Otherwise, it would be sanctioned. Reinforcement learning is commonly used in robot and computer game agent science.
\end{itemize}

\subsection{Optimization objective in machine learning}
All machine-learning algorithms are optimization problems with the goal of evaluating the extremum of an objective function. The development of models and logical objective functions is the first step in machine-learning methods. The determined objective function is normally used with appropriate numerical optimization methods to solve the optimization problem.
\subsubsection{Optimization in Supervised Learning}
The aim of supervised learning is to find an optimal mapping function $F(X)$ that minimises the training samples' loss function.
\begin{equation}\label{equ1}
\begin{aligned}
     F(X) &= \min_{\beta} \frac{1}{N} \sum_{i=1}^{N} l(Y^i,F(X^i,\beta))
\end{aligned}
\end{equation}
where $N$ are training samples, $\beta$ is the mapping function parameter, $X^i$ is $i^th$ samples' feature vector, $Y^i$ is the related label to the data sample, and $l$ is the loss function.

In supervised learning, there are a variety of loss functions, including the square of Euclidean distance, cross-entropy, contrast loss, hinge loss, information gain, and so on. The best way to solve regression problems is to use the square of Euclidean distance as the loss function, reducing square errors on training samples. However, this type of empirical failure does not always work well in terms of generalisation. Structured risk minimization is another popular form, with the support vector machine as the representative system. Regularization items are typically applied to the objective function to prevent overfitting, such as in the $l_2$-norm case.
\begin{equation}\label{equ2}
\begin{aligned}
  \min_{\beta} \frac{1}{N} \sum_{i=1}^{N} l(Y^i,F(X^i,\beta))+\gamma {\vert\vert \beta \vert \vert^2_2}
\end{aligned}
\end{equation}
The compromise parameter, $\gamma$, can be calculated using cross-validation.

\subsubsection{Optimization in Unsupervised Learning}
Clustering algorithms \cite{hartigan1979ak,murtagh1983survey,estivill2000fast,ball1967clustering}, divide data into several classes that are either identical or dissimilar. The k-means clustering algorithm's optimization problem is formulated as minimising the following loss function:
\begin{equation}\label{equ3}
\begin{aligned}
  \min_{D} \sum_{i=1}^{K} \sum_{X\in D_i} \vert\vert X-C_i \vert\vert^2_2
\end{aligned}
\end{equation}
where $K$ denotes the number of clusters, $X$ the sample feature vector, $C_i$ the cluster $i$ center, and $D_i$ the cluster $i$ sample set. The aim of this objective function is to minimise the sum of all cluster variances.

The dimensionality reduction algorithm ensures that the original information from data is retained as much as possible after projecting it into low-dimensional space. A common dimensionality reduction method is principal component analysis (PCA) \cite{wold1987principal,lovric2011international,tipping1999probabilistic}. The goal of PCA is to minimise the reconstruction error as much as possible.
\begin{equation}\label{equ3}
\begin{aligned}
  \min \sum_{i=1}^{N} \vert\vert \Tilde{X}_i-X_i \vert\vert^2_2,\:\:\text{where}\:\Tilde{X}_i &= \sum_{j=1}^{M'} y^i_j.f_j,\:M>>M'
\end{aligned}
\end{equation}
where $N$ are total samples, $X_i$ is a
M-dimensional vector, and $X_i$ is the reconstruction of $X_i$. $y^i$ is the projection of $x_i$ in $M'$-dimensional coordinates. $f_j$ is the standard orthogonal basis under $M'$
-dimensional coordinates.
\subsubsection{Optimization in Reinforcement Learning}
In contrast to supervised and unsupervised learning, RL \cite{mnih2015human,sutton2018reinforcement,kaelbling1996reinforcement} aims to find an optimal strategy function whose performance differs with the environment. The learning goal for a deterministic strategy is the mapping function from state $S$ to action $A$. The learning goal for an unknown strategy is the likelihood of performing each action. $A = \pi(S)$, where $\pi(S)$ is the policy function, determines the behaviour in each state. In RL, the optimization problem can be stated as maximising the cumulative return after performing a series of actions determined by the policy function.
\begin{equation}\label{equ3}
\begin{aligned}
  \max_\pi P_\pi (S),\:\:\text{where}\:P_\pi (S)&=Q \left[\sum_{i=0}^{\infty} \delta^i u_{t+i}|S'_t=S\right]
\end{aligned}
\end{equation}
where, $u$ is the reward, and $\delta \in [0,1]$ is the discount factor and $\pi(S)$ is the value feature of state $S$ under policy $\pi$.

\subsubsection{Optimization in Semisupervised Learning}
SSL is a supervised-unsupervised learning approach that includes both labelled and unlabeled data during the training phase. It can handle a variety of tasks, such as classification \cite{guillaumin2010multimodal,chapelle2005semi}, regression \cite{zhou2005semi}, clustering \cite{demiriz1999semi,kulis2009semi}, and dimensionality reduction \cite{zhang2007semi,chen2017semi}. Self-training, generative models, semisupervised support vector machines (S3VM) \cite{bennett1999semi,sun2019survey}, graph-based methods, multilearning methods, and others are examples of SSL methods. To demonstrate SSL optimization, we use S3VM as an example.
\begin{equation}\label{equ3}
\begin{aligned}
  \min \vert \vert \Gamma \vert \vert R \left [\sum_{i=1}^{l} \eta^i + \sum_{j=l+1}{N} \min(\Delta^i,z^i) \right]\\
  \text{subject to}\:\: Y^i(W.X^i+b)+ \eta^i \geq 1,\eta \geq 0, i=1,...l\\
  W·X_j + b + \Delta^j \geq 1, \Delta \geq 0, j = l + 1,...,N\\
 {-} (W.X^j + b)+ z^j \geq 1, z^j \geq 0
\end{aligned}
\end{equation}
where $C$ is the penalty coefficient, $X$ and $Y$ are the data sample and its label, and $\eta^i$ is the slack variables. If the true label of the unlabeled instance is positive, $\Delta^i$ represents the misclassification error, and $z^j$ represents the misclassification error if the true label is negative.

\newcolumntype{b}{>{\hsize=1.85\hsize}X}
\newcolumntype{a}{>{\hsize=.15\hsize}X}
\begin{table}[htb!]
\centering
\caption{A summary of machine learning types, along with their optimization objectives, advantages, and disadvantages}

\begin{tabularx}{\linewidth}{|>{\centering\arraybackslash}a|>{\centering\arraybackslash}b|>{\centering\arraybackslash}X|>{\centering\arraybackslash}X|}
\hline
Type of ML&Optimisation Function&Advantage&Disadvantage\\\hline
SL&$\min_{\beta} \frac{1}{N} \sum_{i=1}^{N} l(Y^i,F(X^i,\beta))$&It assists in using experience to refine performance criteria, It aids in the solution of a variety of real-world computation problems&Good and numerous examples are needed during training, It takes a lot of computing time to train for supervised learning\\\hline
UL&$\min_{D} \sum_{i=1}^{K} \sum_{X\in D_i} \vert\vert X-C_i \vert\vert^2_2$&It does not require any labeling of data for classification, It is a simple way to reduce the number of dimensions in a dataset&Since we don't have any input data to train from, the outcome could be less accurate, The complexity rises as the number of features grows\\\hline
RL&$\max_\pi P_\pi (S)$&It doesn't necessitate a large number of labelled datasets, This model of learning is remarkably similar to human learning&An excess of states will result from too much reinforcement learning, lowering the quality of the results, It's not recommended to use it to solve basic problems\\\hline
SSL&$\min \vert \vert \Gamma \vert \vert R \left [\sum_{i=1}^{l} \eta^i+\sum_{j=l+1}{N} \min(\Delta^i,z^i) \right]$& Provides little supervision to unlabeled data, It increases efficiency in terms of accuracy&The outcomes of iteration are not consistent, It does not apply to data at the network level\\\hline
\end{tabularx}\label{tab:3}
\end{table}

\newcolumntype{b}{>{\hsize=1.50\hsize}X}
\newcolumntype{a}{>{\hsize=.50\hsize}X}
\begin{table}[htb!]
\centering
\caption{A summary of state-of-art ML-centric resource management  approaches}

\begin{tabularx}{\linewidth}{|>{\centering\arraybackslash}a|>{\centering\arraybackslash}a|>{\centering\arraybackslash}b|>{\centering\arraybackslash}b|}
\hline
Study&Year&Author&Organisation\\\hline
\cite{bianchini2020toward}&2017&Bianchini et al.&Microsoft\\\hline
\cite{haghshenas2020prediction}&2020&Haghshenas et al.&University of Tehran\\\hline
\cite{shaw2019energy}&2019&Shaw et al.&National University of Ireland\\\hline
\cite{9272657}&2021&Ilager et al.&University of Melbourne\\\hline
\cite{nguyen2017virtual}&2017&Heiu et al.&Aalto University\\\hline
\cite{yang2014imeter}&2014&Yang et al.&Beihang University\\\hline
\cite{garg2014sla}&2014&Garg et al.&University of Tasmania\\\hline
\cite{calheiros2014workload}&2014&Calheiros et al.&University of Melbourne\\\hline
\cite{verma2016dynamic}&2016&Verma et al.&University of Hyderabad\\\hline
\cite{subirats2015assessing}&2015&Subirats et al.&Barcelona Supercomputing Centre\\\hline
\cite{messias2016combining}&2016&Messias et al.&University of Sao Paolo\\\hline
\cite{cao2014cpu}&2014&Cao et al.&Shanghai Jiaotong University\\\hline
\cite{shyam2016virtual}&2016&Shyam et al.&Reva Institute of Technology and Management\\\hline
\cite{ismaeel2015using}&2015&Ismaeel et al.&Ryerson University\\\hline
\end{tabularx}\label{tab:3}
\end{table}

\section{Challenges, state-of-art research and their limitations}\label{sec:2}
In this section, we discuss challenges identified in ML-based resource management in state-of-art research. In addition, we explore current approaches to addressing these challenges, as well as their advantages and limitations.

\subsection{Performance and online profiling of workload}
The main components of large commercial providers' workloads are not well addressed in cloud resource management research. For example, they don't look into VMs' lifetime virtual resource consumption. The majority of research focuses on offline workload profiling, which is infeasible because the input workload may not be available until the VMs are not running in production. Online profiling, on the other hand, is challenging because it is difficult to determine when a random VM has exhibited representative behaviour. If the different workload characteristics are accurately predicted with minimal time complexity, resource management can be more effective. As a result, prediction algorithms face another challenge in terms of accuracy and time complexity.

On Microsoft Azure compute fabric, \cite{bianchini2020toward} presented a machine learning-based prediction system. Through a rest API, this system is capable of learning behaviour from historical data and providing predictions online to various resource managers, such as Server health manager, migration manager, Container scheduler, and energy capping manager. They also released detailed Microsoft Azure real-world workload traces from this system, which show that several VMs consistently have peak CPU utilisation in various ranges. In the event of oversubscribed servers, they changed Azure's VM scheduler to use RC benefit predictions. This forecast-based schedule helps to avoid overuse and exhaustion of physical resources. However, (1) they did not consider memory utilisation in released traces or in the predictive system RC, despite the fact that memory utilisation plays a significant role in physical resource exhaustion. (2) They analysed CPU utilisation time series to determine whether a VM is interactive or delay-insensitive, categorised the workload into these two categories, and used Extreme Gradient Boosting Tree (EGBT) to perform supervised classification of these VM workloads. They did not, however, consider the case of a distributed data centre, where data is dispersed and may only have partial labels for these two classes; in this case, there will be insufficient labels to train this algorithm.

\newcolumntype{s}{>{\hsize=.20\hsize}X}
\newcolumntype{a}{>{\hsize=1.40\hsize}X}
\newcolumntype{b}{>{\hsize=1.40\hsize}X}

\begin{landscape}
\begin{table}[ht]
\centering
\caption{A summary of simulation, used datasets and performance improvement of state-of-art research}
\adjustbox{width =\linewidth,totalheight=\textwidth}{
\begin{tabularx}{\linewidth}{|>{\centering\arraybackslash}s|>{\centering\arraybackslash}a|>{\centering\arraybackslash}X|>{\centering\arraybackslash}b|}
\hline
Study&Experiments configuration&Dataset & Performance
improvement \\\hline
\cite{bianchini2020toward}&Online experimentation using real VM traces &Microsoft Azure Trace& Significant prediction accuracies for different workload \\\hline
\cite{haghshenas2020prediction}&Simulation using CloudSim with 7600 hosts& PlanetLab& It reduces the energy consumption upto to 38\% compared to other work. It takes 5\% less time overhead to execute for a modeled data center\\\hline
\cite{shaw2019energy}& Simulation using CloudSim with 800 hosts & PlanetLab& Reduces energy upto 18\% and service violation up to 34\% compared to its baseline\\\hline
\cite{9272657}&Simulation using CloudSim with 75 hosts& Private cloud data from University of Melbourne& Reduces peak temperaturure by 6.5 °C and consumes  34.5\% less energy compared to its baseline\\\hline
\cite{nguyen2017virtual}&\shortstack{Google Cluster
Data\\PlanetLab}&Simulation using CloudSim with 800 hosts& Significantly reduces energy consumption and VM migrations\\\hline
\cite{yang2014imeter}&Simulation using real VM workload&NASA NPB, IOzone and Cachebench& Predicts VM power usage with an average error of 5\% and 4.7\% compared to actual power measurement models\\\hline
\cite{garg2014sla}&Simulation using CloudSim with  1500 physical nodes& Grid Workload Archive (GWA) and PlanetLab&Reduces the number
of servers utilized by 60\% compared to other strategies\\\hline
\cite{calheiros2014workload}& Simulation using CloudSim with 1000 hosts&Wikimedia Foundation& Achieves efficiency in resource utilisation upto 91\%  guaranteeing QoS\\\hline
\cite{verma2016dynamic}&Simulation using two data centers, and three hosts per data center&8 VMs in modeled data centers in CloudSim&Significant allocations of VMs to the host with full capacity\\\hline
\cite{subirats2015assessing}&Experimentation for predictions for different type of workloads&Workloads generated using SPECweb2005& It improves the precision of the forecasts of the energy efficiency while running different workload types
benchmark\\\hline
\cite{messias2016combining}&Experimentation using real web logs&FIFA world cup 98 Web servers, NASA Web servers and ClarkNet Web server&Significant prediction results\\\hline
\cite{cao2014cpu}&Colected CPU load from 12 different hosts&Private cloud environment&Improvement in prediction by 4.81\%, 5.92\% and 7.37\% for BEST MRE, 50\% MREs and 80\% MREs\\\hline
\cite{shyam2016virtual}&Simulation using SamIam Bayesian
network&Amazon EC2 and Google CE data centres& Workload predicted with accuracies greater than 80\%\\\hline
\cite{ismaeel2015using}&Experimentation using real workload VM traces&Google Cluster data&Produces lower RMSE value than other approaches\\\hline
\end{tabularx}}\label{tab:1}
\end{table}
\end{landscape}

\newcolumntype{s}{>{\hsize=.20\hsize}X}
\newcolumntype{a}{>{\hsize=.70\hsize}X}
\newcolumntype{b}{>{\hsize=1.55\hsize}X}
\newcolumntype{f}{>{\hsize=1.55\hsize}X}
\begin{landscape}
\begin{table}[ht]
\centering
\caption{State-of-art research: Objectives, Advantages and Limitations}
\adjustbox{width =\linewidth,totalheight=\textwidth}{
\begin{tabularx}{\linewidth}{|>{\centering\arraybackslash}s|>{\centering\arraybackslash}a|>{\centering\arraybackslash}b|>{\centering\arraybackslash}f|}
\hline
Study&Objectives&Advantages&Limitations\\\hline
\cite{bianchini2020toward}&Online profiling of workload&Predictions are provided online&Memory use is not taken into account, nor is the case of distributed data centres\\\hline
\cite{haghshenas2020prediction}&VM consolidation&Time overhead is considered&Prediction relies on multiple features\\\hline
\cite{shaw2019energy}&VM placement&Dynamic VM placement based on CPU utilisation and network bandwidth&Disc throughput is not considered\\\hline
\cite{9272657}&Thermal management&Peak temperature is reduced significantly&Algorithm overhead\\\hline
\cite{nguyen2017virtual}&VM consolidation based on multiple resource usage&Combination of current and future resource utilization is considered&Overloaded host in the current period of time is not taken into account\\\hline
\cite{yang2014imeter}&Energy consumption prediction&Energy metering at software-level i.e., VM-level&Decision could be taken for an individual VM only based on predicted energy consumption in RMS\\\hline
\cite{garg2014sla}&Resource management strategy based on SLAs&Historical CPU utilisation data with SLA penalties is used&Deviation of prediction from actual value, Highly non-linear workload is not considered\\\hline
\cite{calheiros2014workload}&QoS aware workload prediction&Predicted requests are considered to provision VM dynamically&Future estimation is provided for a static time-interval\\\hline
\cite{verma2016dynamic}&Resource demand prediction and provision strategy&Classification of service tenants based on a binary problem&Information of how binaries are assigned to service tenants is not available, Assigning binaries could be time consuming, Supervised classification could have some limitations in case of partially labelled data\\\hline
\cite{subirats2015assessing}&Energy consumption prediction based on ensemble learning&Ensemble learning is considered&Last-level-cache (LLC), disc throughput are not considered, Accuracy is workload specific\\\hline
\cite{messias2016combining}&Auto-Scaling of web applications&Auto-scaling can adapt to any new workload, Independent of type of prediction models, It can adapt more advanced prediction models&High time complexity\\\hline
\cite{cao2014cpu}&Time-series prediction&Ensemble approach can dynamically adjust the models&Non-generalized approach\\\hline
\cite{shyam2016virtual}&Prediction of virtual resources&Detection of dependencies comprehensively in a variable based on the analysis of non-linear workloads&Combination of several application types is not considered, Non-generalized approach, Transaction throughput and latency are not taken into account\\\hline
\cite{ismaeel2015using}&VM categorization&Use of Extreme learning machine (ELMs), It can deal with non-linear processes, Use of a single network for prediction, Every cluster can have its own network for prediction&Static number of VM clusters.\\\hline
\end{tabularx}}\label{tab:2}
\end{table}
\end{landscape}

\newcolumntype{s}{>{\hsize=.05\hsize}X}
\newcolumntype{b}{>{\hsize=1.0\hsize}X}
\newcolumntype{B}{>{\hsize=1.95\hsize}X}
\begin{landscape}
\begin{table}[htb!]
\centering
\caption{Machine learning-centric resource management challenges and future research directions}
\adjustbox{width =\linewidth,totalheight=\textwidth+.7cm}{
\begin{tabularx}{\linewidth}{|>{\centering\arraybackslash}s|>{\centering\arraybackslash}b|>{\centering\arraybackslash}B|}
\hline
X&Challenges (Section \ref{sec:2}.X)&Future Research Directions (Section \ref{sec:3}.X)\\\hline
1& 
             Online profiling of non linear workload, Prediction accuracy, Time complexity
       &More precise estimate of prior workload using advanced ML models,
       Prediction of memory utilisation in physical resource exhaustion along with CPU utilisation, Semi-supervised classification in categorising VMs\\\hline
2&
            Excessive VM migrations, Host overutilisation, Memory and disc utilisation in VM consolidation
       &Overloaded host detection based on the combination
of CPU, memory, and bandwidth utilisation,
       Workload prediction using DL methods like LSTM, GRU, etc. \\\hline
3&Non-linear resource utilisation,
                     Various resource demands patterns,
                     Cloud network bandwidth
                     &Consideration of disc throughput along with CPU and bandwidth utilisation in VM placement heuristics\\\hline
4&To cool down the host,
                Cost of the cooling system,
                Thermal management
                 & Prior CPU estimation-based resource
provisioning, Use of GRU for inlet temperature prediction only\\\hline
5&
Rapid degradation of load, Unnecessary VM migration overhead&Development of ML algorithm with dynamic resource utilisation threshold\\\hline
6&Prediction of energy consumption at VM level,
Performance degradation due to lower the CPU frequency of server&Prediction of energy consumption state at VM level using clustering analysis\\\hline
7&Resource wastage,
Resource prediction in the presence of computationally
intensive applications&To involve current and future requirements of resources like CPU, memory and bandwidth and SLAs such as compute intensive non-interactive jobs and transactional applications in VM dynamic consolidation, To consider a combination of provisioned and utilised resources like CPU and memory in dynamic resource provisioning\\\hline
8&High utilization,
Exact number of resources in the presence of varying load&To deal with reactive approaches in resource provisioning, Adhoc decisions in dynamic resource provisioning, To predict the peak utilisation of resources using different ML models, Ensemble learning, Estimating future web requests with a dynamic time interval\\\hline
9& To obtain historical data, Amount of resources, Varying resource requirements&To classify service tenants using clustering or semi-supervised clustering
\\\hline
10&Prediction of energy consumption in real time production&To consider memory, disc, and network components system in energy consumption prediction, To inspect non-linear relationships such as polynomial or exponential between virtual resource and energy consumption, Combine information provided by an individual model,  To keep track of the parameters value of each model from the
past record, To feed the ML model with average workload performance for training \\\hline
11&Prediction accuracy in proactive approaches, Limitations of statistical learning over machine learning&To use ML methods to forecast workload instead
of statistical methods, To use feature selection methods such as wrappers, filters, embedded method in ML models\\\hline
12&Arrival of new patters in workload, No single ML model for all time series, Fixed prediction models in ensemble learning approaches&Generalised ensemble framework, Novel models incorporating both global and
local parameters, Ensemble learning, Prediction using advanced neural networks like Temporal Convolution Networks (TCN)\\\hline
13&Dynamic resource prediction, optimal data training in ML model&Optimisation of hyperparameters of ML models using heuristics like Grid Search, Random Search, Bayesian Optimization, Gradient-based
Optimization, and Evolutionary Optimization\\\hline
14&Multiple virtual resources,Demand prediction of each type of virtual resource &To categorize the VMs using advanced clustering approach like cluster ensemble involving clustering accuracy, time complexity and resource usage (CPU and memory utilisation) as model evaluation criteria \\\hline

\end{tabularx}}\label{tab:3}

\end{table}
   % your table

\end{landscape}

\subsection{Multiple Resource Usage in VM Consolidation}
VM consolidation approaches attempt to consolidate more VMs on a smaller number of hosts in order to turn off the remaining hosts and save energy. Most researchers used current CPU utilisation to determine whether a host was overloaded or not in this process. This may result in unnecessary VM migration and host power mode transition, lowering the consolidation process' efficiency. The destination host for migrating VMs is the host with the highest CPU utilisation, but due to the lack of future estimation, this may result in overutilisation. As a result, future resource utilisation estimation can address this issue. Aside from CPU utilisation, other resource consumption, such as memory and disc, can cause the host to become overloaded, making the consolidation process difficult and challenging.

\cite{haghshenas2020prediction} proposed an intelligent VM consolidation technique to reduce energy consumption. Based on historical data, this technique predicted resource utilisation in the past and used that prediction to choose a host with higher utilisation in advance for VM migration. A dynamic consolidation procedure was used to address this issue. To predict the future usage of all VMs, a machine learning method called Linear Regression (LR) was used. This task was carried out using real workload traces from PlanetLab VMs. They used the CloudSim toolkit \cite{calheiros2011cloudsim}  to model a data centre and implement their VM migration strategy to save energy. Their work had the main benefit of taking into account time overheads while lowering energy consumption on a larger simulated benchmark with 7600 hosts. However, if this approach is used in real-world workload production, the time overhead is a significant factor that is also affected by the ML algorithm's data training time. However, they considered the LR method, which relies on various features to predict the target variable, making it time consuming and potentially affecting the data center's response time.

\subsection{Cloud Network Traffic}
The current research in VM allocation involves many solutions to allocate a single VM to a host and allocates various VM resources by ensuring that every host is having sufficient capacity to run the workload. This approach leads to inefficient resource utilisation as the application workload varies time to time with a mix of high and low resource utilisation. The challenges arise when different applications exhibit different resource demands and are allocated to suitable VMs in data centers that cause varying resource demand patterns. Moreover, many VM placement solutions consider only current resource utilisation like CPU demands, however, varying workload continuously poses a challenge to such solutions. Future resources like CPU demand can be more effective for VM placement strategies. In addition to CPU resource demand, cloud network bandwidth is also becoming another challenging factor in efficient resource management in data centers \cite{genez2015estimation}\cite{duggan2017network}. As \cite{networking2016cisco} reported that there will be 51,774GB/sec amount of internet traffic would be produced because of computing as a service via cloud computing and this would affect cloud network as well. And this key factor affects the VM migration time in case of dynamic VM placement and violates SLAs \cite{verma2008pmapper}.

\cite{shaw2019energy} proposed a network-aware predictive VM placement heuristic to reduce energy consumption and SLA violations by considering CPU demand along with the network bandwidth. The main advantage of their work was to design a dynamic VM placement strategy which was based on the prediction of both CPU utilisation and network bandwidth because estimating network bandwidth in case of large VM migration contributes in making decisions with improved scheduling and makes VM placement efficient and reliable. Thus, VM placement strategies should consider future insights of resources to balance limited resource availability and for energy efficient management. However, they did not not consider another aspect, disc throughput, that may also affect VM migration time \cite{brewer2016disks}.

\subsection{Host Temperature}
In modern cloud data centers, minimizing host temperature is a challenging issue. This is caused by the released heat in the process of energy consumption by the host. The cooling systems are deployed to rid of this dissipated heat to keep the host's temperature below the threshold. This increased temperature directly affects the cost of the cooling system and has become a challenging issue to resolve in resource management systems. It also creates host spots in the system and is responsible for several system failures. Thus, thermal management is necessary and challenging due to this dynamic behaviour of the host's temperature.

\cite{9272657} proposed a thermal aware predictive scheduling approach to reduce the peak temperature of a host and energy consumption. Since mostly data centers and servers are having monitoring sensors to record several parameters such as resource usage, energy consumption, thermal reading, and fan speed readings, hence this kind of data was collected from University of Melbourne's private cloud data center. They predicted host temperature by using several machine learning algorithms  and proposed a thermal aware scheduling algorithm to minimize the peak temperature of hosts while migrating VMs to the fewest hosts to reduce energy consumption as well. In this approach, the prediction model is invoked to predict the host temperature and further scheduling is guided. The main advantage of their work is that they reduce the peak temperature up to 6.5$^{\circ}$ and 34\% energy consumption in comparison to existing algorithms, and it was reported by \cite{gao2014machine} that reducing even one degree in temperature can save up to millions of dollars in a large-scale data center. They consider the host's ambient temperature  for prediction instead of CPU temperature that combines inlet temperature and CPU temperature, however, it may increase the algorithm overhead.

\subsection{False Host Overloaded Detection}
The current resource utilisation prediction causes unreliable overloaded host detection, especially in the case of when a current resource utilisation exceeds a threshold value. The challenge arises in deciding whether VMs allocated to this host should be migrated or not because the load decreases rapidly after a very short period of time that leads to a false hot detection point, i.e., false overloaded host detection. However, when the duration of load degradation is large enough, then VMs needs to be migrated to avoid over utilisation. Such kind of VM consolidation mechanism poses a unique challenge to the resource management system to avoid unnecessary VM migration overhead.

\cite{nguyen2017virtual} proposed a VM consolidation strategy based on multiple usage prediction and multi-step prediction for limiting the unnecessary VM migrations to avoid overheads and wasted energy consumption in data centers. Thus, this mechanism was computed to estimate the long-term utilisation of several resources such as CPU, memory based on the historical data for a particular PM. In VM consolidation, the main task it to detect overloaded and underloaded host. Thus, they considered both current and predicted  resource utilisation to identify the overloaded and underloaded hosts. An efficient multiple usage prediction algorithm was presented to compute the long-term utilisation of different resource types based on local historical data. Furthermore, a VM consolidation based on multiple usage prediction was proposed to reduce energy consumption by limiting the  unnecessary VM migrations from overloaded hosts. Hence, the combination of current and predicted resource utilisation plays an important role in reliable overloaded and underloaded host detection. According to this, a host is considered overloaded if it follows two constraints: (1)
if the host is overloaded in both current and predicted resource utilisation, and (2) if the host is in normal condition and will be overloaded in a future period of time. And VM consolidation was performed based on the detected overloaded hosts by following these two constraints. However, they did not consider the case, if a host is overloaded in a current period of time but will not be overloaded in the future period of time, then what about the overloaded host in the current period of time. This point should be considered in VM consolidation scheme.

\subsection{Energy metering at Software-Level}
Modern servers have multiple energy metres to monitor energy usage, but they are unable to monitor the energy of a single virtual machine, which is difficult to do since measuring energy at the software level is difficult. And, according to the energy budget in data centres, energy consumption has become a difficult factor to consider for a successful VM consolidation phase. The previous study only looked at server resource utilisation for VM consolidation, which contradicted the energy capping mechanism by increasing across the levels of certain servers during the process, which violated energy constraints. The term "energy capping" refers to a process introduced at the hardware level. As a result, by lowering the CPU frequency, it reduces the energy consumption of the combined server, which is in violation of the energy constraints. As a result, lowering the server's CPU frequency due to the load of one VM affects all other operating VMs at the same time. As a result, efficiency in workloads running in VMs degrades, breaching SLAs and the isolation property of virtualization. VM consolidation and energy capping are the two most common methods in data centres, but neither allows for accurate monitoring of energy usage for individual VMs.

\cite{yang2014imeter} proposed the iMeter energy consumption prediction model, which is based on the Support Vector Regressor machine learning method (SVR). They used principle component analysis (PCA) to identify the most associated components that influenced VM energy consumption and projected individual VM and multiple consolidated VM energy consumption for various workloads. However, predicting the energy consumption of a single VM is difficult due to the various types of cloud resources residing in the VM, such as CPU, memory, and IO, and the fact that different cloud end users can demand different volumes of the same resources at the same time. Furthermore, the resource manager must make individual decisions for VMs, which slows down end-user response time and violates QoS.

\subsection{SLA-based VM Management}
Over-provisioning has long been used in data centres to prevent the worst-case scenario of peak load utilisation while still meeting SLA obligations. During regular hours, however, the hosts use very few energy, resulting in resource waste. \cite{reiss2012heterogeneity} studied actual workload traces of VMs' resource utilisation from the Google data centre and found that the average CPU and memory utilisation were less than 60\% and 50\%, respectively. Overprovisioning of services, as a result, results in additional maintenance costs in host cooling and administrative activities \cite{sun2016optimizing}. The aim of research has been to solve this difficult problem by using dynamic resource provisioning of resources in virtualization technology, but it primarily focuses on a particular form of SLA or application, such as transactional workload. However, computationally intensive applications are increasingly becoming a part of enterprise data centres, which run multiple types of applications on multiple VMs without taking into account SLA criteria, such as the deadline that results in an under-utilized host. In the case of resource estimation, this factor presents a unique challenge.

\cite{garg2014sla} suggested a novel resource management approach that took into account various types of SLA specifications for various applications operating on various VMs. This approach addresses two types of applications: non-interactive compute-intensive jobs and transactional applications. Both types of applications had a wide range of SLA criteria and specifications. The key benefit of their work was that they used historical CPU utilisation data combined with SLA penalties to forecast potential insight, allowing them to make complex placement decisions in response to shifts in transactional workload and scheduled jobs, taking into account CPU cycles in case of under-utilisation during usual or off-peak periods. The sample of VM CPU usage was used to train an artificial neural network (ANN) to predict VM CPU usage for the next two hours, with the result plotted against actual usage. The X-axis was distributed at a regular interval of 5 minutes. We saw some shortcomings in their work at this point: (1) When there is a wide variance in preparation, the ANN forecast deviates from the actual value in some situations, (2) In a few instances, it also predicts low CPU utilisation from the actual value, (3) They didn't take into account highly non-linear data. The testing data had no non-linear variation, and non-linearity in workload is a major issue nowadays, as data centres have very high non-linearity in workload, which leads to a variety of issues such as high energy consumption, inconsistent QoS, and SLA violations \cite{kumar2020self}.

\subsection{QoS-Aware Resource Provisioning}\label{sec:2.8}
The pattern of evaluating applications deployed on running VMs in modern data centres varies from time to time, i.e., many users attempt to access the application at the same time. As a result, in the cloud, static resource allocation to SaaS applications has been shown to be inefficient because it results in non-linear resource use during periods of low demand and high utilisation. When demand is low, available resources are wasted, resulting in excessive overheads and costs for the cloud service provider; when demand is high, available resources can be inadequate, resulting in weak QoS. This problem can be solved with dynamic resource provisioning, but in this case, the difficulty is determining the correct number of resources to deploy in a given period of time to satisfy QoS requirements when varying workload is available. This challenge is being addressed in two ways: reactively and proactively. The latter has been significantly modified because it is dependent on future load variations prior to their occurrence, i.e., estimating the QoS parameters in advance.

\cite{calheiros2014workload} proposed an ARIMA-based workload prediction model. The main benefit of their work was that the expected requests were used to dynamically provision VMs in an elastic cloud environment while taking into account QoS parameters such as response time and rejection rate. The accuracy of forecast user requests was also assessed in order to see how it affected resource use and QoS parameters. However, we would like to draw your attention to the following limitation in this work. They gathered historical web request data from the Wikimedia Foundation and fed it into a component of their proposed model called \textit{Workload Analyzer}. The ARIMA model was used in this component to provide a future estimation for a specific time interval that can be adjusted for a specific application. The time interval should be long enough to allow for the placement of a new VM for optimal system utilisation. This static time interval may cause issues if a VM deployment time is less than this static time interval, as the extra remaining time may affect QoS parameters such as response time.

\subsection{Varying Patterns of a Service Tenant in Resource Allocation}
Resource demand prediction in a multi-tenant service cloud environment requires historical data to learn the past profiles of service tenants, which is challenging due to the need to update the prediction model on a regular basis because the profiles or trends of service tenants change. AAnother challenge is maintaining the amount of resources required by a service tenant to conduct its operations, which is dependent on many factors, including (1) the operation type, (2) the specific period when the operation is conducted, and (3) the load faced by the service tenant at a specific time. As a result, it presents a challenge because a service tenant's resource requirements can shift. This is a critical topic to address when dealing with resource provisioning using proactive methods for a single service tenant as well as multiple service tenants.

In multi-tenant service clouds, \cite{verma2016dynamic} proposed a dynamic resource demand prediction and provisioning approach to assign resources in advance. They divided the service tenants into groups based on whether or not their resource use would rise in the future. As a result, the proposed system forecast resource demand with priority for only those service tenants whose resource demand was expected to increase, reducing the time required for prediction, which in turn may affect the total time of all operations, thereby affecting QoS. Furthermore, the proposed mechanism used the Best-fit decreasing heuristic method to determine the efficiency of maximum PMs utilisation by combining the service tenants with the matched VMs and allocating them to physical machines (PMs). The most significant aspect of this research is that it classifies service tenants based on a binary issue of whether resource demand will increase or not, and then predicts resource demand for tenants whose resource demand will increase, resulting in a decrease in computational time and cost of prediction. However, (1) we are unable to determine on what basis they mark binaries (0,1) with the service tenants' characteristics, despite the fact that labelling data is needed in order to classify it using supervised learning techniques. (2) If we presume that the service tenants' features were labelled with binaries based on some condition, then labelling the data in a large-scale multi-tenant cloud would be time consuming and would increase the prediction cost. (3) Some data may be accessible without labels in a large-scale distributed multi-tenant cloud, in which case supervised classification would not work.

\subsection{Single ML model in energy consumption prediction}
The majority of cloud service providers' tools calculate and estimate the energy usage of a host or a group of hosts in offline mode, but performing this role in real-time running applications is a challenge. Furthermore, because of the non-linear workload in various hosts, a single ML algorithm cannot be considered to perform this task well. According to \cite{reiss2012heterogeneity}, a Google cluster or node does not use more than 60\% and 50\% of its CPU and memory, respectively. As a result, ensemble learning can be a key component of providing accurate predictions in a cloud architecture. 

\cite{subirats2015assessing} introduced an ensemble learning method for forecasting future energy efficiency in virtual machine resources, such as CPU utilisation, infrastructure, and service levels in a cloud computing environment.Ensemble learning, which uses four different prediction approaches such as moving average, exponential smoothing, linear regression, and double exponential smoothing, is the key benefit of their work.They predict the next use of VM resources, such as CPU consumption, in each time iteration and calculate the mean absolute error (MAE) of all iterations to pick the best performing model predictions for measuring and forecasting energy efficiency and ecological efficiency in an IaaS setting in real time. They do not, however, take into account metrics like Last-level-cache (LLC) and disc throughput for prediction, which have an effect on a host's energy consumption at the VM level \cite{kansal2010virtual}. Furthermore, the accuracy of the chosen model is workload specific, i.e., interactive and batch workloads, rather than being generalised for all data.

\subsection{Prediction Accuracy in Auto-Scaling of web applications}
Auto scaling determines when and how resources are allocated for cloud-based applications. Auto-scaling is done in two ways: reactive and proactive. When system events such as CPU utilisation, number of requests, and queue length exceed a fixed threshold, the reactive approach allocates resources. The proactive approach is in charge of anticipating the amount of resources required ahead of time in order to avoid unneeded events. Furthermore, proactive approaches include predictions based on traditional statistical time-series analysis, which do not fit all cases in terms of prediction accuracy, making it challenging task. Furthermore, statistical learning has the following drawbacks: (1) Statistical Learning is based on rule-based programming, which is formalised as a relationship between variables. (2)Statistical Learning is based on a dataset consisting with a few attributes. (3)Statistical Learning relies on assumptions like normality, no multicollinearity, homoscedasticity, and so on. (4)The majority of the ideas in statistical learning are generated from the sample, population, and hypothesis. (5)Statistical learning is a math-intensive subject that relies on the coefficient estimator and necessitates a thorough knowledge of a dataset.

\cite{messias2016combining} used a genetic algorithm to combine the advantages of individual ML models in order to obtain the best performing prediction results for web application auto-scaling. Each time-series prediction model used in the system is fitted with a suitable weight using a genetic algorithm. The primary benefits of their work are that, (1) Auto-scaling can adapt to any new workload as its characteristics change over time. (2) This approach is unaffected by the type of prediction models used. (3) It's simple to adapt to a variety of more advanced prediction models. However, this approach has a high time complexity, which may affect the response time of any web application hosted in cloud infrastructure, which is in violation of SLAs.

\subsection{Time-Series Prediction Data}
Workload in modern data centres follows a time series pattern. As a result, models for time series prediction should be trained on historical data, as it is presumed that future trends would be identical to those seen previously. However, data centres experience very non-linear workload variations, which is why new trends emerge often, making it difficult for the model to learn precisely. Due to the lack of a single model that is suitable for all types of time series prediction data, an ensemble approach is being used to address this issue \cite{wolski1998dynamically}. Furthermore, most ensemble models for time series prediction are based on a collection of fixed predictors, either homogeneous or heterogeneous, which makes it difficult for the models to learn pattern change in time series prediction.

\cite{cao2014cpu} suggested a new ensemble method that can dynamically update the predictors in the ensemble approach to quickly respond to trend changes in time-series prediction. The ensemble method dynamically adjusts the models, which is the key benefit of this work. It's adaptable, as new models can be quickly added and removed depending on how well it performs with non-linear workload. They set a threshold value of 5 and a floor limit of 0 to determine which predictor is performing well and which is not. Every predictor is given a score, which rises and falls in response to the predictor's results. This predictor is selected as a representative predictor if its score exceeds the threshold value, and it is discarded if it meets the floor limit. These fixed parameters, on the other hand, yield satisfactory results for their chosen dataset, resulting in a non-generalized approach.

\subsection{Data Training}
In modern cloud environments, virtual resources such as virtual CPUs (vCPUs) and memory (vRAMs) have a non-linear resource demand, resulting in complex resource utilisation behaviour. As a result, with this high amount of workload on a daily basis, optimization of virtual resource performance is required. Large corporations such as Amazon, Alibaba, and others have occasionally failed due to a lack of resource management planning. As a result, predicting virtual resources (such as vCPU and vRAM) is a challenging task. Furthermore, resource forecasting presents some challenges, (1) The prediction of these resources should be dynamic in order to respond to changing workload patterns over time, (2) The data for training should be chosen in such a way that it has the greatest impact on the target variable, so that the model can learn to predict it effectively.

\cite{shyam2016virtual} proposed a model that took into account a variety of parameters in a virtualized platform to reliably predict virtual resources with the least amount of SLA violations. This method was based on a Bayesian approach that identified various variables and took into account the best training data. The key benefit of their work is that it detects dependencies in a variable in a systematic manner based on the study of non-linear workloads from various data centres such as Amazon, EC2, and Google. However, (1) they do not take into account the combination of several application types, (2) Since it relies on the dependencies of a specific problem, this approach lacks generalisation, (3) For prediction, this method ignores high-level metrics including transaction throughput and latency of underlying resources, such as vCPU cores.

\subsection{VM Multi Resources}\label{sec:2.14}
Flexible resource provisioning frameworks are needed in cloud data centres to manage host load based on various requirements. As a result, data centres conduct dynamic resource provisioning, which uses prediction models to estimate the amount of resources needed in advance for varying workloads over time. Its aim is to predict future VM request workloads by looking at previous usage trends. However, since VM requests include a variety of virtual resources such as CPU, memory, disc, and network throughput, it is extremely challenging and complex to forecast demand for each form of resource separately. In the case of choosing an ML prediction model, the multi-resource existence of a VM presents a specific challenge. Furthermore, different cloud users can make different requests for cloud resources. As a result, forecasting the demand for each form of resource is difficult and impractical.

\cite{ismaeel2015using} proposed a model for dividing VM clusters into different categories and then developing prediction models for each cluster. The key benefit of their work is that (1) they use Extreme Learning Machines (ELMs), which can find the best weight for the predictor in a single step. (2) They avoid issues like stopping conditions, learning rate selection, learning epoch scale, and local minimums of gradient-based learning methods like NN and ANFIS by using ELMs. (3)As it deals with non-linear processes, this work can handle the linear behaviour of the LR method. (4) It predicts VM requests in each cluster using a single network. (5) Every cluster can have its own prediction network. However, in kmeans clustering, they set the number of clusters to 3, resulting in a model with a fixed number of VM clusters.

\section{Future Research Directions}\label{sec:3}
\subsection{Performance and Online Profiling of Workload}
The efficiency of the intelligent resource management system is determined by many factors, including the accuracy and time complexity of the prediction model. Huge corporations such as Google, Microsoft, Amazon, and others are in charge of extremely complex data centres with a wide range of workloads. As a result, in the presence of such a highly variable or nonlinear workload for VMs, a more accurate estimation of prior workload is a future research direction by employing more sophisticated ML and DL modes. Furthermore, the time complexity of an algorithm is a measurement of its performance in terms of the time it takes to run the input code. As a result, the algorithm should be designed to be as simple as possible in terms of time complexity. Furthermore, online profiling is necessary to prevent VM blackouts until they are running in development, as well as various resource utilisation such as CPU and memory, which are major contributors to physical resource exhaustion and should be considered for prediction. \cite{cortez2017resource,bianchini2020toward} conducted online workload profiling and provided an analysis to determine if a virtual machine is interactive or delay-insensitive. To categorise VMs into these two groups, they used supervised classification. In this situation, semi-supervised learning \cite{zhu2009introduction} may play a vital role and may be a potential research direction to train the data with these partial labels and perform classification with promising accuracy in large-scale distributed data centres.

\subsection{Multiple Resource Usage in VM consolidation}
A host is considered overloaded during the VM consolidation phase if CPU utilisation reaches a throughput threshold, such as 80\% \cite{nguyen2017virtual}. However, other resource utilisation, such as memory use and bandwidth use \cite{abdelsamea2017virtual}, leads to host overloading. As a result, detecting overloaded hosts using a combination of CPU, memory, and bandwidth use is a potential research direction in the VM consolidation phase. For an efficient VM consolidation operation, the estimation of current and future CPU, memory, and bandwidth use should be addressed. The current study \cite{abdelsamea2017virtual,haghshenas2020prediction} involves a variety of machine learning algorithms, such as linear regression and multiple regression, in which the model's training is based on multiple features in order to simulate a target variable, such as CPU utilisation. The training time of multiple features will affect the VM migration time in the VM consolidation process, which affects QoS and SLAs in large-scale distributed data centres where millions of VMs are running in production. As a result, dealing with the training time of ML models is a potential future research direction. Different deep learning (DL) approaches, such as Long Short-Term Memory (LSTM) networks \cite{hochreiter1997long} and Gated Recurrent Unit (GRU) \cite{cho2014learning}, can deal with training time by avoiding the overheads of multiple features by using a single feature, such as a vector of CPU utilisation, as an input for training to predict its next state in the future.

\subsection{Cloud Network Traffic}
The problem of varying patterns of various types of workloads when considering current resource utilisation in VM allocation on a host is a challenge. As a result, predicting potential resource demand, such as CPU and network bandwidth, has proven to be an alternative approach \cite{shaw2019energy}.However, in addition to these resources, disc throughput is a significant factor to consider. In VM placement heuristics, taking disc throughput into account is a new research direction. It calculates the amount of data that can be stored, read, and written per second. \cite{brewer2016disks} published a report stating that disc tail latency, especially reads, is a key factor when delivering online services where a user is waiting for a response.As a result, disc throughput can play a role in VM migration time, affecting tail latency time and violating SLAs. Therefore, according to our vision, a prior maximum estimate of disc throughput will play a critical role in avoiding delay.

\subsection{Host Temperature}
\cite{9272657} proposed a scheduling algorithm to minimise the host temperature that was driven by the host temperature prediction computed using several ML algorithms. As a consequence, estimating host temperature ahead of time can help with thermal management decisions like VM migration to reduce host temperature, i.e., CPU temperature. \cite{9272657}, on the other hand, took into account the ambient temperature for prediction, which is a combination of CPU and inlet temperature. This could result in an increase in algorithm overhead. Furthermore, they discovered that the host's CPU temperature is primarily affected by CPU load and power consumption. As a result, it is being waited for the CPU to become overloaded, causing the temperature to rise, resulting in additional cooling costs for the host. As a potential future research topic, Prior CPU estimation-based resource provisioning can prevent the CPU from becoming overloaded and save energy. Then we'll only have to deal with the inlet temperature, which may reduce the thermal management algorithm's overhead. Furthermore, several ML algorithms necessitate a significant amount of training time due to the training of multiple features, which can slow down VM migration. It will cause VM migration to be delayed, which will slow down host temperature degradation and add to the cost. Thus, using an ML or DL method like GRU, where the inlet temperature can be used as an input to train a model that can predict its future state using single feature training, could be an alternative. Doing so can avoid an overhead algorithm, a delay in VM migration, a delay in minimising the host temperature.

\subsection{False Host Overloaded Detection}
The overloaded host detection's static threshold can result in unreliable VM migration. If the utilisation of a VM's resources degrades in a short period of time, there is no need to migrate the VM. In this case, the algorithm should have a dynamic resource utilisation threshold that automatically prevents VM migration when it reaches the fixed threshold, taking into account near-future data. For efficient VM migration in VM consolidation, this is the future research direction. Furthermore, VMs should be migrated if the near future information has a long period of load degradation.

\subsection{Energy metering at Software-Level}
Many power management decisions, such as power capping, will benefit from visibility of energy usage at the host and VM levels. At the host level, energy consumption is simple to predict or calculate since modern data centres have several built-in sensors that track it, but it is difficult to measure at the VM level because to measure the energy consumption induced by memory, we must collect LLC (last-level-cache) events raised by each VM on each core, which is difficult to do \cite{kansal2010virtual,zhao2012power}. Rather than calculating or predicting energy consumption at the VM level, clustering analysis may be used to determine the status of VMs in terms of energy consumption, such as low, moderate, or critical. Thus, dividing VMs by conducting clustering analysis based on highly co-related features with energy consumption at the VM-level is a potential research direction, and there would be no need to obtain host-level features. ML techniques such as ChiSquare Score, Fisher Score, Gini Index, and Correlation-based Feature Selection (CFS) can be used to find the correlation with energy consumption \cite{vora2017comprehensive}. Then, using a clustering algorithm or a clustering ensemble \cite{kadhim2019rapid}, a clustering analysis can be performed to determine which VMs are in low and critical energy consuming states. By doing so, a group of VMs can be managed together in a data center's resource management system, potentially reducing response time and improving QoS.

\subsection{SLA-based VM Management}
Future research directions for avoiding non-linear resource utilisation in modern data centres include dynamic resource provisioning and dynamic VM consolidation, which take into account various types of VM resources such as CPU, memory, and bandwidth, current and future resource needs, and SLAs such as compute intensive non-interactive jobs and transactional applications. Both of these methods rely heavily on accurate resource prediction. \cite{garg2014sla}, for example, provided long-term CPU utilisation forecasts that differed significantly from actual test phase data due to a substantial shift in CPU utilisation during the training phase, which is critical for dealing with non-linear utilisation in modern data centres. Future research will focus on optimising hyper parameters used in Artificial Neural Network (ANN) learning, such as mini batchsize, epochs, and number of neurons. The model is said to work better if it is trained on the data in an optimised manner. The observation of the validation and loss graphs estimated with these optimised hyperparameters may indicate that the model has learned a lot when both plots begin moving closely and consistently, and learning should be stopped at these optimised parameters.

\subsection{QoS-Aware Resource Provisioning}
The aim of this study is to use constructive dynamic resource provisioning based on workload estimation using historical data to improve QoS parameters like response time and rejection rate. Future research could concentrate on dealing with it in a reactive manner, with resource provisioning occurring after resource demand, such as the number of requests, has arrived. Furthermore, according to the current study \cite{calheiros2014workload}, the error in request prediction can be mitigated by adhoc decisions in dynamic resource provisioning, which can help to boost poor QoS efficiency. Furthermore, there is a potential research direction to forecast peak CPU use using more sophisticated ML models such as XGBoost \cite{chen2016xgboost}, LSTM \cite{hochreiter1997long}, and GRU \cite{cho2014learning} in a correct manner that cannot be equipped with the ARIMA model.  Furthermore, no single machine learning algorithm can suit any non-linear workload with time-series data, necessitating an ensemble learning approach in which various ML and DL methods can be used in the future. After that, the best-performing model can be selected for potential use. \cite{calheiros2014workload}, as discussed in Section \ref{sec:2.8}, estimates web requests based on a static time interval that can affect response time. As a result, it can be addressed by estimating future web requests with a dynamic time interval that adjusts automatically based on the VM deployment time. In such a way that the time interval of estimation can be equivalent to the VM deployment time and the remaining time can be avoided if the VM deployment time is much shorter than this static time interval that affects the QoS parameter as the response time. Prior estimation of VM deployment time based on historical data should therefore be computed and used in the above-mentioned case to satisfy the condition of equivalence with the estimated time of the request prediction.

\subsection{Varying Patterns of a Service Tenant in Resource Allocation}
Clustering analysis, which does not require any data labelling, could be used to classify service tenants as a future research direction. On the basis of historical resource demands, similar patterns of service tenants can be automatically obtained. By observing the similarity between data using clustering, service tenants with high and low resource demand can be distinguished, and predictions for those with high resource demand can be provided using ML and DL regression techniques. In the case of a distributed data centre where data is dispersed and partial labels are available, a concept known as semi-supervised clustering \cite{smieja2020classification} can be used, in which unsupervised data is given a little supervision using partial labels and techniques such as instance-level constraints \cite{wagstaff2000clustering} and relative distance constraints \cite{cho2014learning}.

\subsection{Single ML model in energy consumption prediction}
Apart from the CPU, a system power model includes memory, disc, and network components, so these components could be considered as well. The current study looks at the linear relationship between these metrics and energy consumption; however, non-linear relationships, such as polynomial or exponential, could be explored in the future. In addition, in an ensemble learning approach, the best individual model is chosen, which may or may not be the best solution. Another option is to combine the information provided by each individual model and analyse the results. This can be accomplished by estimating the average using weights based on each individual predictor's mean average error. Furthermore, each workload type requires its own set of configuration parameters. The future research direction is to keep track of the parameters value of each model from the past record that have increased the maximum utilisation of resources and to use them in real-time scenarios to adapt the models to the workload type of each individual VM. In addition, the forecast accuracy is also affected by a sudden change in the use of resources. A further future research direction is therefore to feed the ML model with average workload performance, such as CPU utilisation.

\subsection{Prediction Accuracy in Auto-Scaling of web applications}
Machine learning models, rather than statistical methods, may be used to predict workload in the future, which has many advantages: (1) Machine Learning learns from data without the need for explicit programming. (2) Machine Learning has the ability to learn from billions of observations and features, (3) Machine Learning relies less on assumptions and, in most cases, disregards them. (4) Machine Learning emphasizes predictions, supervised learning, unsupervised learning, and semi-supervised learning (5) Machine Learning uses iterations to identify patterns in a dataset, requiring far less human effort. The training of multiple features is needed to predict the target variable, which increases the time complexity of machine learning methods like regression. As a result of the existence of redundant features, ML methods suffer from latency and computational complexity problems when processing multiple features. In such datasets, the number of functions, feature dependency, number of records, feature types, and nested feature categories all substantially increase the processing time of ML methods. As a result, future research should concentrate on using suitable feature selection methods, such as wrappers, filters, embedded methods, and enhanced versions \cite{majeed2019improving}, to effectively overcome the computation speed versus accuracy trade-off when processing large and complex datasets.

\subsection{Time-Series Prediction Data}
The development of a generalised ensemble framework for any type of dataset in cloud time series workload data is a future research direction. Deep learning (DL), in general, is a rapidly expanding and broad research field that involves novel architectures. However, researchers are never sure when they need to adapt which methods to which situations. \cite{hewamalage2021recurrent} used global NN models, which are prone to outlier errors in some time series. As a result, novel models incorporating both global and local parameters for individual time series must be developed in the form of hierarchical models. These models can be combined with ensembling, which involves training multiple models with the same dataset in different ways. Furthermore, CNNs have long been used for image processing, but they are now being used to forecast time series data. According to \cite{lai2018modeling,shih2019temporal}, traditional RNN models are ineffective at modelling seasonality in time series forecasting. As a result, they combine CNN filters for local dependencies and a custom attention score function for long-term dependencies. In order to capture seasonality patterns, \cite{lai2018modeling} has also tried recurrent skip connections. \cite{oord2016wavenet} developed Dilated Causal Convolutions to effectively capture long-range dependencies along the temporal dimension. They've recently been used in conjunction with CNNs to solve problems involving time series forecasting. Temporal Convolution Networks (TCN), which combine dilated convolutions and residual skip connections, have also been introduced as more advanced CNNs \cite{borovykh2017conditional}. According to \cite{bai2018empirical} TCNs are promising NN architectures for sequence modelling tasks, in addition to being efficient in training. As a result, using CNNs instead of RNNs could provide a competitive advantage for forecasting practitioners. As a consequence, these potentially advanced neural networks could be used in the future to forecast workload time series in cloud infrastructure.

\subsection{Data Training}
The aim of optimising machine learning hyperparameters is to find the hyperparameters for a particular machine learning algorithm that achieves the best performances on validation data. The hyperparameters are set by the engineer before the training, contrary to the model parameters. The number of trees in a random forest, for example, is a hyperparameter, whereas the weights in a neural network are model parameters learned during training. Size and decay are support vector machine hyperparameters (SVM) and k in k-nearest neighbours (KNN), respectively. Furthermore, hyperparameter optimization returns an optimal model that reduces a predefined loss function and, as a result, improves the accuracy on given independent data by finding a combination of hyperparameters. Hyperparameters can thus have a direct effect on machine learning algorithm training. It is therefore critical to understand how to optimise them in order to achieve maximum performance. This points to a future research direction of optimising the hyperparameters of ML algorithms for achieving optimal dataset training. This can be accomplished by employing some common heuristics such as Grid Search, Random Search, Bayesian Optimization, Gradient-based Optimization, and Evolutionary Optimization \cite{feurer2019hyperparameter}.

\subsection{VM Multi Resources}
As stated in Section \ref{sec:2.14}, there is a future research direction to categorize the VMs and develop a prediction model for each cluster to address the multi-resource demand challenges. However, the use of a clustering algorithm such as kmeans can limit the number of clusters available, causing a VM to be placed in the incorrect cluster. A clustering ensemble can be a better approach than clustering because it aims to combine multiple clustering algorithms to produce a final consensus solution that is more robust and accurate than a single clustering algorithm \cite{alqurashi2019clustering}. This literature \cite{boongoen2018cluster} mentions a number of clustering ensemble methods. Furthermore, in a recent work \cite{kadhim2019rapid}, two additional evaluation criteria such as time complexity and resource usage (CPU and memory usage) were considered to evaluate the novel clustering ensemble, in addition to clustering accuracy. Thus, advanced clustering methods such as clustering ensemble can be used in the future to achieve the best clusters with the highest precision, least time complexity, and least resource consumption. 

\section{SUMMARY AND CONCLUSIONS}
In this paper, we discuss the challenges of machine-learning-based resource management in a cloud computing environment, as well as the various approaches that have been used to solve these challenges in recent years, along with their benefits and drawbacks. In recent years, there has been a significant increase in the number of studies looking at how to use machine learning techniques to conduct workload prediction, energy consumption prediction, and other tasks. Different ML methods are used in these techniques to deal with various types of problems. Finally, based on the challenges and drawbacks identified in the state-of-the-art work, new potential future research directions are proposed to strengthen the current ML methods for resource management in cloud-based systems. The overall knowledge provided in this paper aids cloud researchers in comprehending cloud resource management and the significance of machine learning techniques.

Our findings show that machine learning models can be used in cloud computing systems to achieve various optimization goals and deal with complex tasks. The use of ML approaches also opens up a new avenue for intelligent resource and application management. This article illustrates the progress of machine learning approaches in current research and helps readers understand the research gap in this field.
To improve system efficiency, one promising way is to use advanced machine learning techniques such as reinforcement learning and deep learning to perform intelligent resource management.
.

\bibliographystyle{unsrt}
%\bibliography{bibiliography}  
%\bibliographystyle{ACM-Reference-Format}
\bibliography{main}
\end{document}